\documentstyle[camera,prbplug,psfig]{article}

\begin{document}
\draft
\title{The ground state of Sr$_3$Ru$_2$O$_7$ revisited; Fermi liquid
 close to a ferromagnetic instability }
\author{Shin-Ichi Ikeda$^{1,2}$, Yoshiteru Maeno$^{2,3}$, Satoru 
Nakatsuji$^{2}$, 
Masashi Kosaka$^4$, and Yoshiya Uwatoko$^4$}
\address{$^1$Electrotechnical Laboratory, Tsukuba, Ibaraki 305-8568
, Japan}
\address{$^2$Department of Physics, Kyoto University, Kyoto 606-8502,
 Japan}
\address{$^3$CREST, Japan Science and Technology Corporation, Kawaguchi, 
Saitama 332-0012, Japan}
\address{$^4$Department of Physics, Saitama University, Saitama 338-8570, 
Japan}

\maketitle
\begin{abstract}
We show that single-crystalline Sr$_3$Ru$_2$O$_7$ grown by a floating-zone 
technique is an isotropic paramagnet and a quasi-two dimensional metal 
as spin-triplet superconducting Sr$_2$RuO$_4$ is. The ground state is 
a Fermi liquid with very low residual resistivity ($\approx $3 $\mu \Omega 
$cm for in-plane currents) and a nearly ferromagnetic metal with the largest 
Wilson ratio $R_{\rm W} \ge 10$ among paramagnets so far. This 
contrasts with the ferromagnetic order at $T_{\rm C}$=104 K reported 
on single crystals grown by a flux method [Cao {\it et al.}, Phys. Rev. B 
$\bf{55}$, R672 (1997)]. However, we have found a dramatic changeover 
from
 paramagnetism to ferromagnetism under applied pressure. This suggests 
the existence of a substantial ferromagnetic instability in the Fermi liquid state.

\end{abstract}

The discovery of superconductivity in the single-layered perovskite 
Sr$_2$RuO$_4$ \cite{maeno1} has motivated the search for new 
superconductors and anomalous metallic materials in Ruddlesden-Popper 
(R-P) type ruthenates (Sr,Ca)$_{n+1}$Ru$_n$O$_{3n+1}$. 
 The recent determination of the spin-triplet pairing in its superconducting 
state suggests that ferromagnetic (FM) correlations are quite important in 
Sr$_2$RuO$_4$ \cite{ishida1}, and the existence of enhanced spin fluctuations 
has been suggested by nuclear magnetic resonance (NMR) 
\cite{mukuda1,imai1}. On the other hand, the recent report has shown that 
enhanced magnetic excitations around $\bf{q}$=0 is not detected but 
sizeable excitations have been seen around finite $\bf{q}$ in Sr$_2$RuO$_4$ 
by inelastic neutron scattering\cite{sidis}. This has stimulated debate on 
the mechanism of the spin-triplet superconductivity, which had been naively 
believed to have a close relation to FM ($\bf{q}$=0) spin excitations. Hence,
 it is desirable to investigate its related compounds as described below.

The simple perovskite (three dimensional) metallic SrRuO$_3$ ($n= \infty $) 
has been well known to order ferromagnetically below 160 K with a magnetic 
moment $M=0.8 \sim 1.0 \mu_{\rm B}/$Ru \cite{kanb1,kiyama1}. FM 
perovskite oxides are relatively rare except for metallic manganites. For pure 
thin film SrRuO$_3$, analyses of quantum oscillations in the resistivity 
have given good evidence for the Fermi liquid behavior \cite{andy}. 

The double layered perovskite Sr$_3$Ru$_2$O$_7$ ($n$=2) is regarded as 
having an intermediate dimensionality between the systems with $n=1$ and 
$n=\infty$ \cite{will1}. Investigations on polycrystalline 
Sr$_3$Ru$_2$O$_7$ showed a magnetic-susceptibility maximum 
around 15 K with Curie-Weiss-like behavior above 100 K and a metallic 
temperature dependence of the electrical resistivity 
\cite{cava1,ikeda1}.

In the study presented here, we have for the first time succeeded in growing 
single crystals of Sr$_3$Ru$_2$O$_7$ by a floating-zone (FZ) method. 
Those single crystals (FZ crystals) do not contain any impurity phases 
(e.g. SrRuO$_3$) which was observed in polycrystals \cite{ikeda1}. 
 We report herein that the FZ crystal of Sr$_3$Ru$_2$O$_7$ is a nearly
FM paramagnet (enhanced paramagnet) and a quasi-two dimensional
 metal with a strongly-correlated Fermi liquid state. In addition, we have 
performed magnetization measurements under hydrostatic pressures up to 
1.1 GPa in order to confirm whether the FM instability is susceptible to 
pressure. The results suggest that there is a changeover from paramagnetism 
to ferromagnetism, indicating a strong FM instability. Essential features 
of magnetism for FZ crystals as well as polycrystals are inconsistent
with the appearance of a FM ordering ($T_{\rm c} = 104 {\rm K}$) at
 ambient pressure for 
single crystals grown by a flux method \cite{cao1} using SrCl$_2$ flux and Pt 
crucibles. We will argue that FZ 
crystals reflect the intrinsic behavior of Sr$_3$Ru$_2$O$_7$.  

Details of the FZ crystal growth are explained elsewhere \cite{ikeda2}. 
The crystal structure of the samples at room temperature was characterized 
by powder x-ray diffraction. Electrical resistivity $\rho (T)$ was measured 
by a standard four terminal dc-technique from 4.2 K to 300 K and by an ac 
method from 0.3K to 5K. Specific heat $C_{P}(T)$ was measured by a relaxation
method from 1.8 K to 35 K (Quantum Design, PPMS). The temperature 
dependence of magnetic 
susceptibility $\chi (T) \equiv M/H$ from 2 K to 320 K was measured 
using a commercial SQUID magnetometer (Quantum Design,  
MPMS-5S). For magnetization measurements of FZ crystals at ambient 
pressure, we performed sample rotation around the horizontal axis, 
normal to the scan direction, using the rotator in MPMS-5S. We could align
 the 
crystal axes exactly parallel to a field direction within 0.2 degree using this 
technique. For high pressures, we measured magnetization using a long-type 
hydrostatic pressure micro-cell \cite{uwa} with the SQUID 
magnetometer. Loaded pressures around 3 K were determined from the shift 
of superconducting transition temperature of Sn in the micro-cell in a 5 mT field . 

The R-P type structure of $n$=2 for FZ crystals of Sr$_3$Ru$_2$O$_7$ 
was confirmed by the powder x-ray diffraction patterns with crushed crystals, 
which indicated no impurity peaks. Recently, the crystal structure of 
polycrystalline Sr$_3$Ru$_2$O$_7$ has been refined by neutron powder 
diffraction \cite{huang,shaked}. Although they have concluded that symmetry 
of the structure is orthorhombic owing to the rotation of the RuO$_6$ octahedron 
about the c-axis by about 7 degrees, we deduced lattice parameters at room 
temperature by assuming tetragonal $I4/mmm$ symmetry as 
$a= 3.8872(4) {\rm \AA}$, and $c=20.732(3) {\rm \AA}$. These values 
are in good agreement with those of polycrystals obtained by neutron diffraction 
\cite{huang,shaked} and x-ray diffraction \cite{ikeda1}. 

The temperature dependence of magnetic susceptibility $\chi (T)=M/H$ in
 a field of 0.3 T is shown in Fig. 1. No hysteresis is observed between zero-field 
cooling (ZFC) and field cooling (FC) sequences, so we conclude that there is no 
ferromagnetic ordering. Little magnetic anisotropy is observed in contrast to large 
anisotropy ($\approx 10^2$) of flux-grown crystals \cite{cao1}. The nearly 
isotropic susceptibility of Sr$_3$Ru$_2$O$_7$ is qualitatively similar to that of 
the enhanced Pauli-paramagnetic susceptibility in Sr$_2$RuO$_4$ 
\cite{maeno2}.
 For an applied field of 0.3 T, there is no in-plane anisotropy of the susceptibility 
for the whole temperature range (2 K $\le T \le $ 300K), within the precision
 of our equipment (1$\%$). 

\begin{figure} 
\centerline{\psfig{file=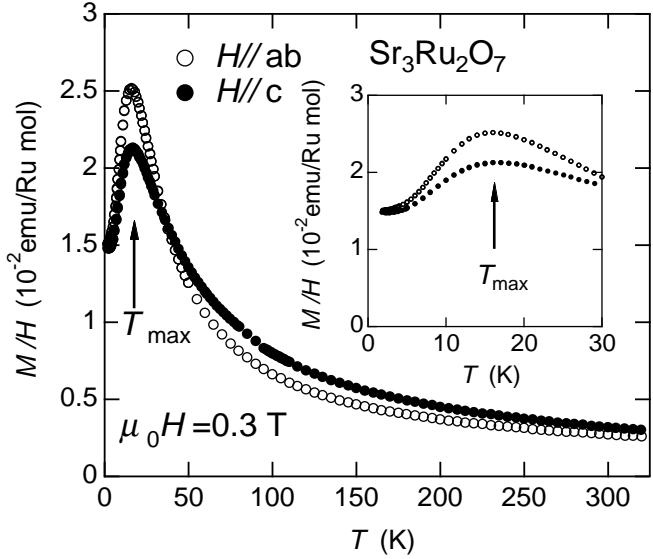,width=\columnwidth}}
\caption{Magnetic susceptibility of FZ crystals of 
Sr$_3$Ru$_2$O$_7$ under 0.3 T field above 2 K. 
The inset shows the low temperature magnetic susceptibility against 
temperature $T$.} 
\end{figure} 

 The susceptibility for both $H//$ab and $H//$c exhibits Curie-Weiss behavior 
above 200 K. We have fitted the observed $\chi (T)$  from 200 K to 320 K 
with $\chi (T)=\chi_{\rm 0}+\chi_{\rm CW}(T)$, where $\chi_{\rm 0}$ 
is the temperature independent term and $\chi_{\rm CW}(T) = 
C/(T-{\it \Theta}_{\rm W})$ is the Curie-Weiss term. The effective Bohr 
magneton numbers $p_{\rm eff}$ deduced from $C$ are $p_{\rm eff}=
$ 2.52 (2.99) and ${\it \Theta}_{\rm W}$ =$\relbar $ 39 K ($\relbar 
$ 45 K) for $H//ab (H//c)$. The negative values of ${\it \Theta}_{\rm W}$ 
normally indicate antiferromagnetic (AFM) correlations in the case of 
localized-spin systems. However, we cannot conclude that AFM 
correlations play an important role solely by the negative ${\it \Theta}_
{\rm W}$ in an metallic system like Sr$_3$Ru$_2$O$_7$ \cite{yoshimura}.  

Around $T_{\rm max}$ =16 K, $\chi (T)$ shows a maximum for both 
$H//$ab and $H//$c. The maximum has been also observed in the 
polycrystals. The results of temperature dependence of specific heat, NMR 
\cite{mukuda} and elastic neutron scattering \cite{huang,shaked} for 
polycrystals indicate that there is no evidence for any long range order 
with definite moments. The FZ crystal shows nearly isotropic $\chi 
(T)$ for all crystal axes below $T_{\rm max}$. Hence, the maximum 
cannot be accredited to the long range AFM order. Therefore, we conclude 
Sr$_3$Ru$_2$O$_7$ to be a $paramagnet$. Concerning $\chi (T)$ under higher
 fields, 
$T_{\rm max}$ is suppressed down to temperatures below 5 K above 6 T
\cite{perry}. 
Such a maximum in $\chi (T)$ and a field dependent $T_{\rm max}$ are
 often observed in a nearly ferromagnetic (enhanced paramagnetic) metal like 
TiBe$_2$ \cite{jarl} or Pd \cite{muell}. In addition, a similar 
behavior in $\chi (T)$ has been observed in (Ca,Sr)$_2$RuO$_4$ 
\cite{c3po1} and MnSi \cite{pfle}, which are recognized as examples 
of a critical behavior by spin fluctuations. Similar critical behavior, 
originating especially from FM spin fluctuations, is also expected in 
Sr$_3$Ru$_2$O$_7$. Nevertheless, we cannot rule out the possibility 
of AFM correlations as observed in Sr$_2$RuO$_4$, caused by the 
nesting of its Fermi surfaces with the vector $\bf{Q}$ $= 
(\pm 0.6 \pi/a, \pm 0.6 \pi/a,0)$ \cite{sidis}.

As shown in Fig.2, the specific heat coefficient of the FZ crystal of 
Sr$_3$Ru$_2$O$_7$ is $\gamma $ = 110 mJ/(K$^2$ Ru mol) 
somewhat larger compared to other R-P type 
ruthenates \{$\gamma = $80 mJ/(K$^2$ Ru mol) for CaRuO$_3$, 30 
mJ/(K$^2$ Ru mol) for SrRuO$_3$ \cite{kiyama1} and 38 
mJ/(K$^2$ Ru mol) for Sr$_2$RuO$_4$ \cite{maeno2,andy1}\}.
 This suggests that Sr$_3$Ru$_2$O$_7$ is a strongly-correlated 
 metallic oxide. For polycrystalline Sr$_3$Ru$_2$O$_7$, we obtained 
the value $\gamma = 63 $ mJ/(K$^2$ Ru mol) using an adiabatic method 
\cite{ikeda1}.

\begin{figure} 
\centerline{\psfig{file=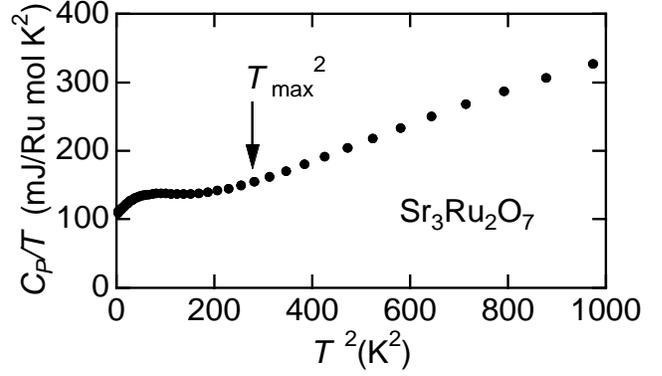,width=\columnwidth}}
\caption{Specific heat devided by temperature $C_P/T$ of FZ crystals of 
Sr$_3$Ru$_2$O$_7$ above 2 K. $C_P/T$ is plotted against $T^2$.} 
\end{figure} 

The temperature dependence of the electrical resistivity $\rho (T)$ is shown 
in Fig. 3 above 0.3 K. Both $\rho_{\rm ab}(T)$ and $\rho_{\rm c}(T)$
 are metallic ($d\rho/dT >0$) in the whole region. The ratio of 
$\rho_c$/$\rho_{ab}$ is about 300 at 0.3 K and 40 at 300 K. This anisotropic 
resistivity is consistent with the quasi-two-dimensional Fermi surface sheets 
obtained from the band-structure calculations \cite{hase1}. With lowering 
temperature below 100 K, a remarkable decrease of $\rho_{\rm c}(T)$ is 
observed around 50 K. This is probably due to the suppression of the thermal 
scattering with decreasing temperature between quasi-particles and phonons as 
observed in Sr$_2$RuO$_4$ \cite{yoshida,maeno2,andy2}. Thus, below 50K, 
interlayer hopping propagations 
of the quasi-particle overcome the thermal scattering with phonons. This hopping 
picture for $\rho_{\rm c}(T)$ is well consistent with the large value of 
$\rho_{\rm c}(T)$ and nearly cylindrical Fermi surfaces.
On the other hand, $\rho_{\rm ab}(T)$ shows a 
change of the slope around 20 K.  Such a change in $\rho _{\rm ab}(T)$ has 
also been reported for Sr$_2$RuO$_4$ under hydrostatic pressure 
($\approx$ 3 GPa). That might be possibly due to the enhancement of 
ferromagnetic spin fluctuations \cite{yoshida}.

\begin{figure} 
\centerline{\psfig{file=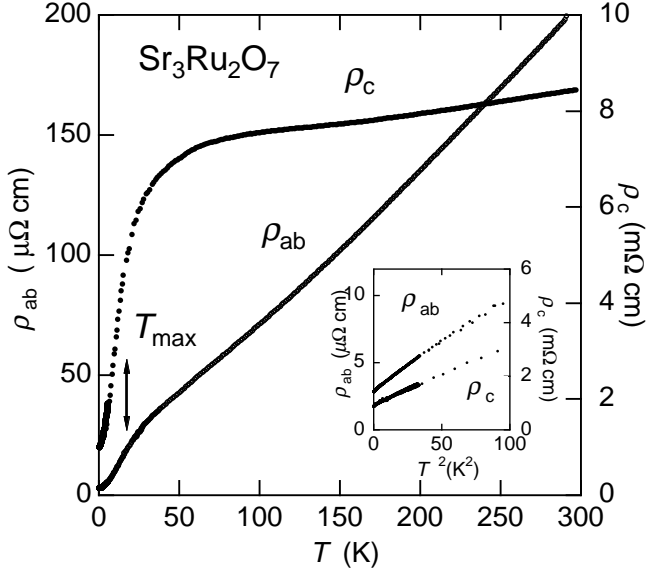,width=\columnwidth}}
\caption{Electrical resistivity of FZ crystals of Sr$_3$Ru$_2$O$_7$ 
above 0.3 K. Both $\rho _{\rm ab}$ and $\rho _{\rm c}$ are shown.
The inset shows the low temperature electrical resistivity against the square
 of temperature $T^{2}$.} 
\end{figure} 

As shown in the inset of Fig. 3, the resistivity yields a quadratic
 temperature dependence below 6 K for both $\rho_{\rm ab}(T)$ and 
$\rho_{\rm c}(T)$ , characteristic of a Fermi liquid as observed in 
Sr$_2$RuO$_4$ \cite{maeno2}. We fitted $\rho_{\rm ab}(T)$ by the 
formula $\rho_{\rm ab}(T)$=$\rho_{\rm 0}$+$AT^2$ below 6 K and 
obtained $\rho_{\rm 0}$=2.8 $\mu \Omega$ cm and $A$=0.075 $\mu 
\Omega $cm/K$^2$. Since the susceptibility is quite isotropic and 
temperature independent below 6 K, the ground state of 
Sr$_3$Ru$_2$O$_7$ 
is ascribable as a Fermi liquid. We now can estimate Kadowaki-Woods ratio 
$A/\gamma ^2$. Assuming that electronic specific heat $\gamma $ 
= 110 mJ/(K$^2$ Ru mol) is mainly due to the ab-plane component, we 
obtain $A/\gamma ^{2} \approx A_{\rm ab}/\gamma ^{2} = 0.6 
\times 10^{\relbar 5} \mu \Omega$ cm/(mJ/K$^{2}$ Ru mol)$^{2}$ 
close to that observed in heavy fermion compounds. 

      Regarding to $\chi(T)$ again, it is important to note that even at 
temperatures much lower than $T_{\rm max}$, $\chi (T)$ remains 
quite large. It appears that the ground state maintains a highly enhanced 
value of $1.5 \times 10^{-2}$ emu/Ru mol, comparable to that 
obtained for typical heavy fermion compounds. Considering that the 
observed $\chi$ is dominated by the renormalized quasi-particles, we 
can estimate the Wilson ratio $R_{\rm W} = 7.3 \times 10^{4} \times 
 \chi ({\rm emu/mol})\slash \gamma ({\rm mJ/(K^{2} mol})$) 
in the ground state. If we regard the observed values at $T$ = 2 K 
as that at $T$ = 0 K, we have $R_{\rm W} =10 (18)$ using $\gamma$ 
for single crystals (polycrystals). Despite the difference in the $\gamma$ 
value between polycrystals and single crystals, $R_{\rm W}$ is much 
greater than unity. This large value implies that FM correlations are 
strongly enhanced in this compound, especially when compared with 
the values of 12 for TiBe$_2$ and 6 for Pd \cite{julian}. Therefore, the 
ground state of Sr$_3$Ru$_2$O$_7$ is characterized by strongly-correlated 
Fermi liquid behavior with enhanced FM spin fluctuations, 
i.e. Sr$_3$Ru$_2$O$_7$ is a strongly-correlated nearly FM metal.

Concerning this FM correlations, it should be noted that, using single 
crystals grown by a chlorine flux method with Pt crucibles \cite{muller}, Cao 
{\it et al.} have investigated remarkable magnetic and transport 
properties of R-P ruthenates \cite{cao1} prior to our crystal growth. 
The ground state of Sr$_3$Ru$_2$O$_7$ was concluded to be an itinerant 
{\it ferromagnet} with $T_{\rm c} = 104$ ${\rm K}$ and an ordered 
moment $M=1.2$ $\mu_{\rm B}/$Ru. The flux-grown crystals were 
reported to have a residual resistivity ($\rho_0=3$ m$\Omega $ cm ) 
$10^3$ times greater than that of FZ crystals ($\rho_0=3 \mu 
\Omega $ cm ) for in-plane transport. In addition, FZ crystals
reveal $T$-square dependent resistivities at low temperatures as 
already shown, which was not observed in flux-grown crystals. In general,
 the FZ method with great care can be impurity-free crystal growth, while
 the flux method tends to contaminate crystals due to impurity elements 
from both the flux and the crucible. This might be a main reason why 
the resistivity is much higher for flux-grown crystals. Thus, we suppose 
with assurance that the data from FZ crystals reflect the intrinsic nature of 
Sr$_3$Ru$_2$O$_7$ better than those from flux-grown crystals.

In order to acquire the information of the magnetic instability in the FZ crystal 
of Sr$_3$Ru$_2$O$_7$, we have measured magnetization under hydrostatic 
pressure up to 1.1 GPa. The temperature dependence of magnetization 
$M(T)$ is shown for several pressures under a 0.1 T field along c-axis in Fig. 4.
 Around 1 GPa, substantial increase is recognized below around 70 K with a 
clear FM component indicated by the difference between ZFC and FC 
sequences. Although the remanent moment at 2 K ($M\approx 0.08$ 
$\mu_{\rm B}/$Ru) is much smaller than that expected for $S$=1 
of Ru$^{4+}$, its susceptibility is quite large (0.4 emu/Ru mol). We infer 
that this transition is a FM ordering of itinerant Ru$^{4+}$ spins. In Fig. 4, we 
also show the field dependence 
of magnetization $M(H$//c) at 2 K for $P=0.1$ MPa and $P=1$ GPa. Obvious 
ferromagnetic component appears at lower fields for $P=1$ GPa. Even at higher 
fields, increase in magnetization by pressure is also present as at lower fields.
 This feature endorses the drastic changeover from paramagnetism to 
ferromagnetism induced by applied pressure. 
To the best of our knowledge, this is the first example of the pressure-
induced changeover from Fermi liquid to ferromagnetism.

\begin{figure} 
\centerline{\psfig{file=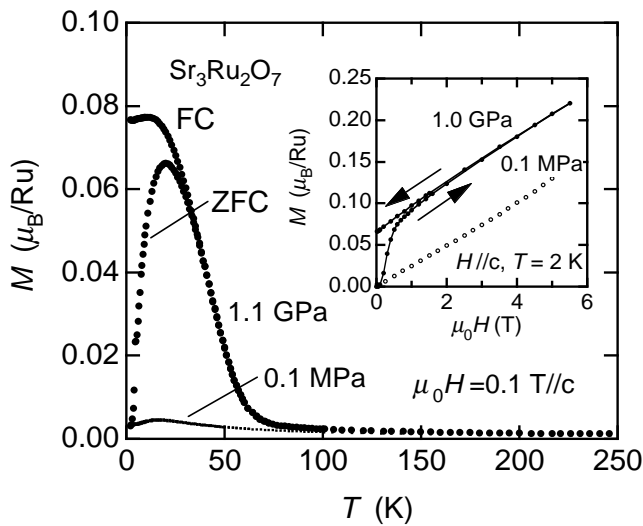,width=\columnwidth}}
\caption{Pressure dependence of magnetization $M(T)$ for $H//$c. 
Obvious ferromagnetic ordering appears at about 70 K under 1 GPa 
pressure. The inset shows the field dependence of magnetization $M(H)$ 
under 0.1 MPa and 1 GPa pressures.} 
\end{figure}

For the purpose of understanding the observed behavior, we should begin 
with Stoner theory. In the metallic state with correlated electrons, the 
ferromagnetic order is driven by the Stoner criterion 
$U_{\rm eff}N(E_{\rm F})\geq 1$, where $U_{\rm eff}$ is an 
effective Coulomb repulsion energy.  The systematics of band-width 
$W$ and the density of states $N(E_{\rm F}$) in the R-P ruthenates 
is summarized by Maeno $et$ $al$ \cite{maeno3}. In this system, 
increasing $n$ from 1 to $\infty$ causes enhancement of 
$N(E_{\rm F}$) as well as $W$. This is opposite to the single band 
picture, i.e. increasing $N(E_{\rm F})$ naively means decreasing $W$. 
In the case of R-P ruthenates, the anomalous variation might be due to 
the modifications of the degeneracy of three $t_{2g}$ orbitals for Ru-$4d$ 
electrons. According to the summary \cite{maeno3}, ferromagnetic 
SrRuO$_3$ is characterized by the highest $N(E_{\rm F}$) and $W$ 
among them, satisfying the Stoner criterion.  This implies that the 
enlargement of $N(E_{\rm F}$) and $W$ reflects stronger three 
dimensionality in the R-P ruthenates. Hence, applying 
pressure probably makes Sr$_3$Ru$_2$O$_7$ closer 
to SrRuO$_3$, leading to FM order. For further investigations, it is 
required that structural study, resistivity and specific heat under 
 pressures will be performed.

In conclusion, by using the floating-zone method we have succeeded for 
the first time in growing single crystals of Sr$_3$Ru$_2$O$_7$ with very 
low residual resistivity in comparison with that of flux-grown crystals
 reported previously. The results of magnetization, resistivity and 
specific heat measurements suggest that Sr$_3$Ru$_2$O$_7$ is 
a strongly-correlated Fermi liquid with a nearly ferromagnetic 
ground state, consistent with the observation of ferromagnetic ordering 
below 70 K under applied pressure ($P \sim $1 GPa). As far as we 
know, this is the first example of the pressure-induced changeover 
from Fermi liquid to ferromagnetism. This ferromagnetic ordering may 
guarantee the existence of the ferromagnetic spin fluctuations in 
Sr$_3$Ru$_2$O$_7$. 

Authors are very grateful to A. P. Mackenzie for his fruitful advice and critical 
reading of this manuscript. They thank T. Ishiguro, T. Fujita and K. Matsushige 
for their helpful supports. They thank S. R. Julian, G.G. Lonzarich, G. Mori, 
D.M. Forsythe, R.S. Perry, K. Yamada, Y. Takahashi, and M. Sigrist for their 
useful discussions and technical supports. They also thank N. Shirakawa for 
his careful reading of this manuscript.


\end{document}